\documentclass[twocolumn,times]{aastex62}

\usepackage{amsmath,amstext}

\shorttitle{The radio emission of AT2018cqh}

\newcommand{\erg}{${\rm erg \ s^{-1}}$ }

\def\ltsima{$\; \buildrel < \over \sim \;$}
\def\simlt{\lower.5ex\hbox{\ltsima}}
\def\gtsima{$\; \buildrel > \over \sim \;$}
\def\simgt{\lower.5ex\hbox{\gtsima}}
\newcommand{\msun}{{\rm\,M$_\odot$}}

\newcommand{\srcs}{{\rm\,AT2018cqh}}
\newcommand{\src}{{\rm\,AT2018cqh }}

\begin{document}

\title{
Delayed and fast rising radio flares from an optical and X-ray detected tidal disruption event in the center of a dwarf galaxy 
}
\correspondingauthor{Xinwen~Shu} 
\email{xwshu@ahnu.edu.cn}

\author{Fabao Zhang}
\affil{Department of Physics, Anhui Normal University, Wuhu, Anhui, 241002, China}

\author[0000-0002-7020-4290]{Xinwen~Shu}
\affil{Department of Physics, Anhui Normal University, Wuhu, Anhui, 241002, China}

\author{Lei Yang}
\affil{Department of Physics, Anhui Normal University, Wuhu, Anhui, 241002, China}


\author{Luming Sun}
\affil{Department of Physics, Anhui Normal University, Wuhu, Anhui, 241002, China} 

\author{Zhumao Zhang}
\affil{Department of Physics, Anhui Normal University, Wuhu, Anhui, 241002, China} 

\author{Yibo Wang}
\affil{Department of Astronomy, University of Science and Technology of China, Hefei, Anhui 230026, China}

\author{Guobin~Mou}
\affil{Department of Physics and Institute of Theoretical Physics, Nanjing Normal University, Nanjing 210023, China. }

\author{Xue-Guang Zhang}
\affil{Guangxi Key Laboratory for Relativistic Astrophysics, School of Physical Science and Technology, GuangXi University, Nanning, 530004, China}

\author{Tianyao~Zhou }
\affil{Department of Physics, Anhui Normal University, Wuhu, Anhui, 241002, China}

\author{Fangkun~Peng }
\affil{Department of Physics, Anhui Normal University, Wuhu, Anhui, 241002, China}

\begin{abstract}
\src is a unique tidal disruption event (TDE) discovered in a dwarf galaxy.  
Both the light curve fitting and galaxy scaling relationships suggest a central black hole mass in the range of $5.9<~$log$M_{\rm BH}/M_{\sun}<6.4$.  
A delayed X-ray brightening was found around 590 days after the optical discovery, 
but shows unusual long-time rising to peak over at least 558 days,  
which could be coming from delayed accretion of a newly forming debris disk. 
We report the discovery of delayed radio flares around 1105 days since its discovery, 
characterized by an initial steep rise of $\simgt$175 days, a flattening lasting about 544 days, 
and a phase with another steep rise.  
The rapid rise in radio flux coupled with the slow decay in the X-ray emission 
points to a delayed launching of outflow, perhaps due to a transition in the accretion state. 
However, known accretion models can hardly explain the origins of the secondary radio flare that is rising even more rapidly 
in comparison with the initial one. 
If confirmed, \src would be a rare TDE in a dwarf galaxy exhibiting optical, X-ray and radio flares. 
We call for continued multi-frequency radio observations to monitor its spectral and temporal evolution, 
which may help to reveal new physical processes that are not included in standard TDE models. 
\end{abstract}


\section{Introduction} \label{sec:intro}
It is now widely accepted that most, if not
all, massive bulge-dominated galaxies harbor supermassive black holes
(SMBH, $M_{\rm BH}\simgt10^{6}$\msun) in their nuclei \citep[e.g.,][]{McConnell2013}.  
A population of intermediate-mass black holes (IMBHs, $M_{\rm BH}\sim10^{4}-10^{6}$\msun) likely exists, and may 
live in the centers of dwarf galaxies \citep[$M_{\star}\simlt10^{9}$\msun,][]{Greene2012, Reines2013}. 
However, identifying BHs in dwarf galaxies and 
measuring their masses (if present) {are} not trivial, as they are typically faint and 
the gravitational influence is expected to be small \citep[e.g.,][]{Reines2022}. 
When a star passes too close to an SMBH, it
can be squeezed and torn apart once the tidal force of the
SMBH exceeds the star’s self-gravity \citep{Stone2019}. 
Such tidal disruption events {(TDEs)} can generate luminous flares typically peaking in X-rays and ultraviolet, 
as a fraction of the disrupted stellar debris falls back and {gets} accreted by the BH \citep{Rees1988, Gezari2021},  
providing a direct way to probe the SMBHs in otherwise quiescent galaxies \citep{Mockler2019}. 

The first TDE candidates were identified as soft X-ray outbursts by the ROSAT all-sky Survey, 
which are thought to connect with thermal emission from a newly formed accretion disk \citep[e.g., ][]{Bade1996, Saxton2020}. 
Thanks to the development in wide-field survey capabilities, especially those at optical bands, 
dozens of TDEs have been identified, making it possible the demographics studies such as correlations between {light curve} properties, 
host galaxies, volumetric rates, and luminosity function \citep{vanVelzen2021, Hammerstein2023, Yao2023}. 
While most optically-selected TDEs are faint in X-rays within the first few months of discovery, 
some {show} the late-time X-ray brightening \citep[e.g., ][]{Gezari2017, Shu2020, Guolo2023}, which 
can be attributed to the delayed onset of accretion \citep{Piran2015} or 
ionization break out of X-ray radiation from the obscuration by 
optically thick outflows in the early phase \citep{Metzger2016}. 
With a flux-limited TDE sample discovered by the Zwicky Transient Facility \citep[ZTF,][]{Bellm2019} over 3 years, \citet{Yao2023} inferred 
a flat BH mass function in the regime of $10^{5.3}\simlt\rm(M_{BH}/M\sun)\simlt10^{7.3}$, 
indicating that IMBHs can be revealed with TDEs. 
Particularly, the short rise time and fast evolution from TDEs in dwarf galaxies are possibly 
the signatures of IMBHs \citep{Angus2022}. 

While the growing number of TDEs is discovered in wide-field optical surveys, 
rapid follow-up observations have led only a few detections \citep{Alexander2020}. 
The radio detection rate appears to increase with observations on longer {timescale} of years since 
the discovery \citep{Cendes2023}.  
Several scenarios have been proposed to explain this late-time radio brightening, 
including a delayed launching of the outflow compared to the time of debris fall-back \citep{Horesh2021a}, 
decelerating of an off-axis jet launched at the time of disruption \citep[e.g.,][]{Giannios2011, Matsumoto2023}, 
propagation of the outflow in an inhomogeneous medium \citep[e.g.,][]{Nakar2007}, 
{outflow--cloud interaction \citep{mou2021,mou2022}},
and break out of the choked precessing jets from the accretion disk wind \citep{Lu2023, Teboul2023}. 
Discriminating between these models will help to determine dependence of jet production on 
the accretion rate, and/or diagnose the density and its radial structure of circumnuclear medium (CNM). 

SRGe J023346.8-010129 was reported as a candidate TDE in a dwarf 
galaxy ($z=0.048$) by SRG/eROSITA \citep{Bykov2023} on 2019 Nov 13. 
Prior to the X-ray discovery, the object has also been named as an optical transient by Gaia Alerts Team (Gaia18bod) on 2018 June 16, but was not classified. 
We will use its TNS ID of AT2018cqh as the transient's name throughout. 
If the optical transient records the time of tidal disruption, this suggests a delay of X-ray brightening 
by $\sim$590 days.  
Here we report the detection of delayed and rapidly rising radio emission from \srcs,  
about 1105 days after the optical alert. 
The observations and data reductions are described in Section 2. 
In Section 3, we present the detailed analysis of optical {light curves}, radio flux and SED evolution properties. 
Discussion on the origins of delayed radio and X-ray emission from \src is given in Section 4. 
We summarize the results in Section 5.
We adopt a cosmology of $\Omega_M$ = 0.3, $\Omega_{\lambda}$ = 0.7, and $H_0$ = 70 km s$^{-1}$ Mpc$^{-1}$ when computing luminosity distance.

\begin{deluxetable*}{cccccc}
  \centering
\tablewidth{0pt}
\tablehead{
\colhead{Observatory} & \colhead{Project} & \colhead{$\nu$} & \colhead{Date} & \colhead{Phase} & \colhead{F$_\nu$} \\
\colhead{} & \colhead{} & \colhead{(GHz)} & \colhead{} & \colhead{(days)} & \colhead{(mJy/beam)}}
\tablecaption{Summary of the radio observations of AT2018cqh \label{tab:table}}
\setlength{\tabcolsep}{3mm}{
\startdata
VLA & VLASS1 & 3.0 & 2017 Nov 30 & -198 & $\textless$ 0.45$^{\dag}$\\
 & VLASS2 & 3.0 & 2020 Sep 26 & 833 & $\textless$ 0.51$^{\dag}$\\
 & VLASS3 & 3.0 & 2023 Mar 07 & 1725 & 10.580 $\pm$ 0.280 \\
\hline
ASKAP & VAST & 1.36 & 2021 Nov 19 & 1252 & 2.590 $\pm$ 0.110 \\
\hline
ASKAP & VAST & 0.89 & 2019 Aug 27$\sim$2020 Aug 28 & 437$\sim$804 & $\textless$ 0.364$^{\ddag}$\\
 & EMU & 0.94 & 2021 Jun 25 & 1105 & 1.037 $\pm$ 0.014 \\
 & EMU & 0.94 & 2021 Nov 07 & 1240 & 2.223 $\pm$ 0.032 \\
 & FLASH & 0.86 & 2021 Dec 17 & 1280 & 2.723 $\pm$ 0.038 \\
 & VAST & 0.89 & 2023 Jun 14 & 1814 & 3.663 $\pm$ 0.093 \\
 & VAST & 0.89 & 2023 Jul 06 & 1846 & 3.744 $\pm$ 0.076 \\
 & FLASH & 0.86 & 2023 Aug 11 & 1882 & 7.132 $\pm$ 0.050 \\
 & VAST & 0.89 & 2023 Aug 30 & 1901 & 8.110 $\pm$ 0.180 \\
 & VAST & 0.89 & 2023 Oct 29 & 1961 & 9.010 $\pm$ 0.170 \\
\enddata}
\tablecomments{
$^{\dag}$For VLASS non-detections, the corresponding 3$\sigma$ upper limits on peak flux density are given. \\
$^{\ddag}$The upper limit was measured by stacking the VAST images observed between 2019 Aug 27 and 2020 Aug 28.  
}
\end{deluxetable*}

\section{OBSERVATION AND DATA} \label{sec:style}

\subsection{Optical photometric and spectral data}

\begin{figure}[htbp!]
\epsscale{1.15}
\plotone{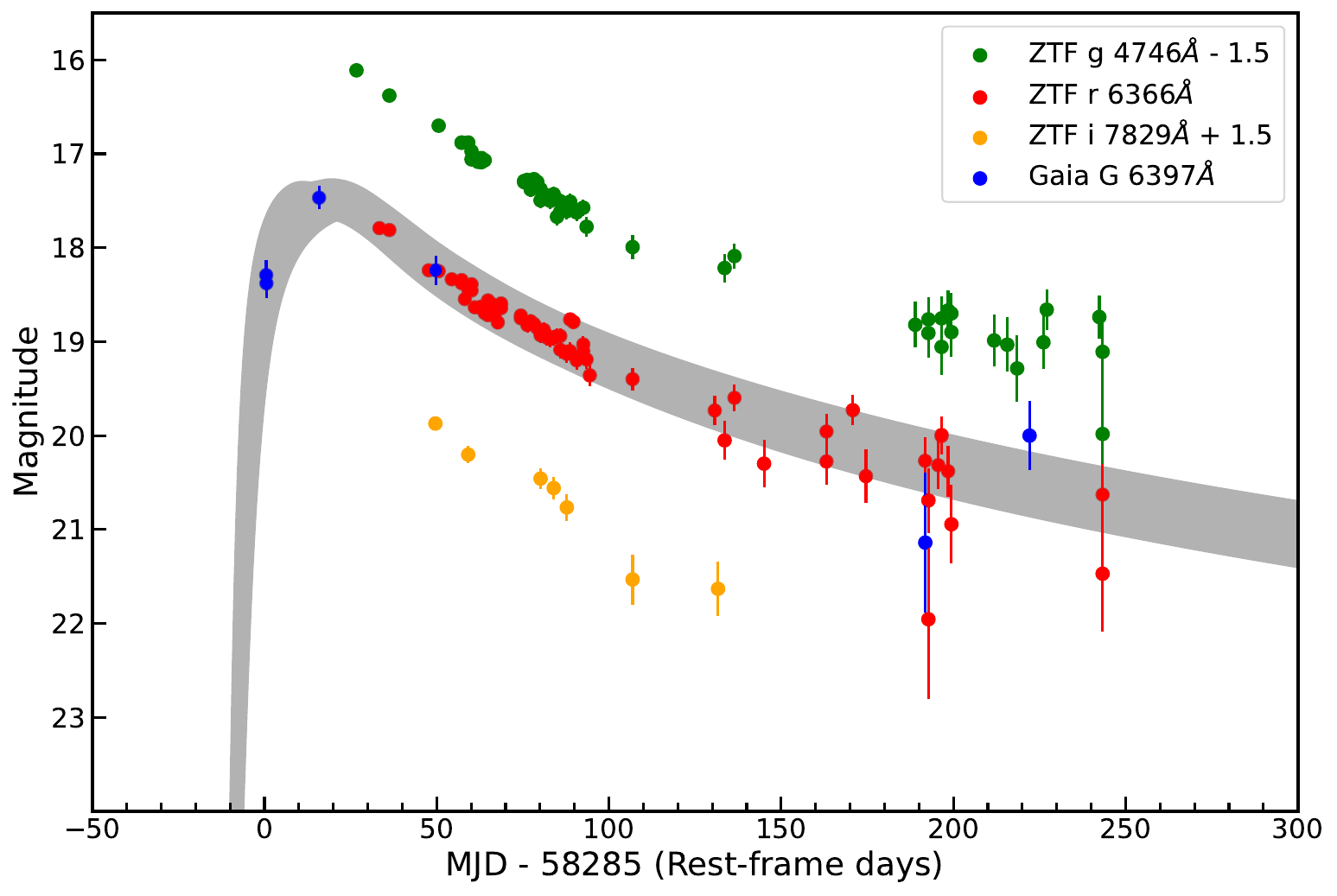}
\caption{
Optical light curves of \src observed with ZTF and Gaia. 
{Note that the quiescent host emission has been subtracted in all light curves.}
We performed joint fits to the ZTF r-band and Gaia G-band light curves with a TDE model using the {\tt MOSFiT} code. The best model realizations are shown in gray curves, which are constructed from the posterior parameter distribution at the 68\% confidence level.
\label{fig:optlc}}
\end{figure}

As shown in Figure \ref{fig:optlc}, we collected the optical light curves of \src obtained by ZTF\footnote{https://ztf.snad.space/dr17/view/401310100001492} and Gaia (from its Alerts website\footnote{http://gsaweb.ast.cam.ac.uk/alerts/alert/Gaia18bod/}).   
ZTF {detects} a prominent flare at both its g-, r- and i-bands, followed by a fading back to the baseline level in $\approx$200 days \citep[see also, ][]{Bykov2023}. 
{While the rising phase was missed by ZTF observations, it was caught by Gaia observations.}
The first flux rising epoch (MJD=58285) was set to be the discovery time of optical flare, which is on 2018 June 16.  
{Using this discovery time and the occurrence time of the X-ray brightening from \citet{Bykov2023}, we estimated a more precise time delay between the optical and the X-ray flares to be $\approx$590 days.}
We measure an offset between the transient position reported in the ZTF observations {during the optical flare (RA = $02^{\rm h}33^{\rm m}46\fs9308$ and DEC = -01$\arcdeg$01$\arcmin$28\farcs3009)\footnote{https://alerce.online/object/ZTF18abtgunq} 
and the host optical centroid reported by the {\it Gaia} DR3 \citep[RA = $02^{\rm h}33^{\rm m}46\fs9339$ and DEC = -01$\arcdeg$01$\arcmin$28\farcs3742,][]{Gaia2021}}, which is 87.1 milliarcseconds\footnote{Since \src is an extragalactic galaxy, we used the {\it Gaia} position without taking into account proper motions.}. 
This corresponds to a physical offset of 82.5 pc at the redshift of \srcs, making it consistent with a nuclear origin. 
We also examined the light curves of \src from the Asteroid Terrestrial Impact Last Alert System (ATLAS), 
which confirm the optical flare but the sampling is sparse. 
Hence, we do not consider the ATLAS data in the following analysis.

\begin{figure*}[ht!]
    \epsscale{1.15}
\plotone{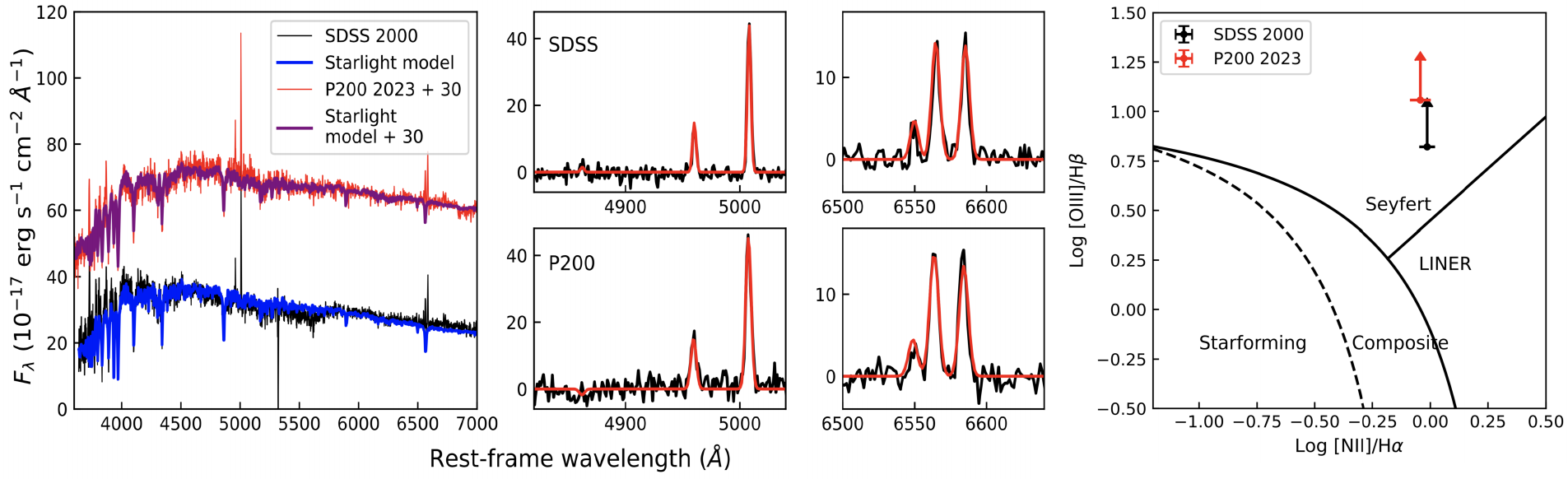}
\caption{{\it Left panel}: P200/DBSP spectrum observed {on} 2023 Oct 6, as compared with the archival SDSS spectrum taken {on} 2000 Oct 3. The corresponding stellar continuum models are also shown (purple and blue). {\it Middle panel}: a zoomed-in view of the emission-line profile fittings for the $H_\beta$, {\sc [O iii]}, and $H_\alpha$+{\sc [N ii]} doublet lines. Gaussian line models are plotted in red. {\it Right panel}: The optical classification of \src in the BPT diagram based on the emission line ratios of {\sc [O iii]}/$H\beta$ vs. {\sc [N ii]}/$H\alpha$. The optical spectrum displays Seyfert-like narrow emission-line ratios which have not changed between P200 and SDSS observations. 
\label{fig:optspec}}
\end{figure*}

Two optical spectra were acquired for \srcs, including archival one from
SDSS DR7 (observed {on} 2000 Oct 03) and one from our own spectroscopic follow-up with 
Double Beam Spectrograph (DBSP) on the 200-inch Hale telescope at Palomar Observatory (\citealt{Oke1982}), which was taken {on} 2023 Oct 6, 1938 days since the optical discovery. 
We used the D55 dichroic which splits the incoming photons into the 600/4000 (lines/mm) grating for the blue side, and 316/7500 grating for the red side. 
The grating angles were adjusted to achieve a nearly continuous wavelength coverage from 3300 to 10000 \AA~.
We reduced the P200/DBSP spectrum with the python package Pypeit (\citealt{Pypeit1,Pypeit2}), which can highly automatically implement the standard reduction procedure for long-slit spectroscopic observations.
Figure \ref{fig:optspec} shows the post-flare P200 spectrum as compared with the pre-flare SDSS spectrum of the host.
{We modelled the two spectra with the python package BADASS (\citealt{Sexton2021}) and the best-fitting models are also shown in the figure.}
No significant change in the spectral features, e.g., continuum and emission lines, 
is observed before and after the transient event, indicating that {either} any optical signatures had faded by the time of our follow-up observation, or there is no associated emission-line echo. 
The latter scenario seems consistent with non-detection of mid-infrared echo from the WISE light curve \citep{Bykov2023}. 
 
\subsection{Radio data} 

\begin{figure*}[tph!]
  \begin{center}
\includegraphics[width=0.93\textwidth]{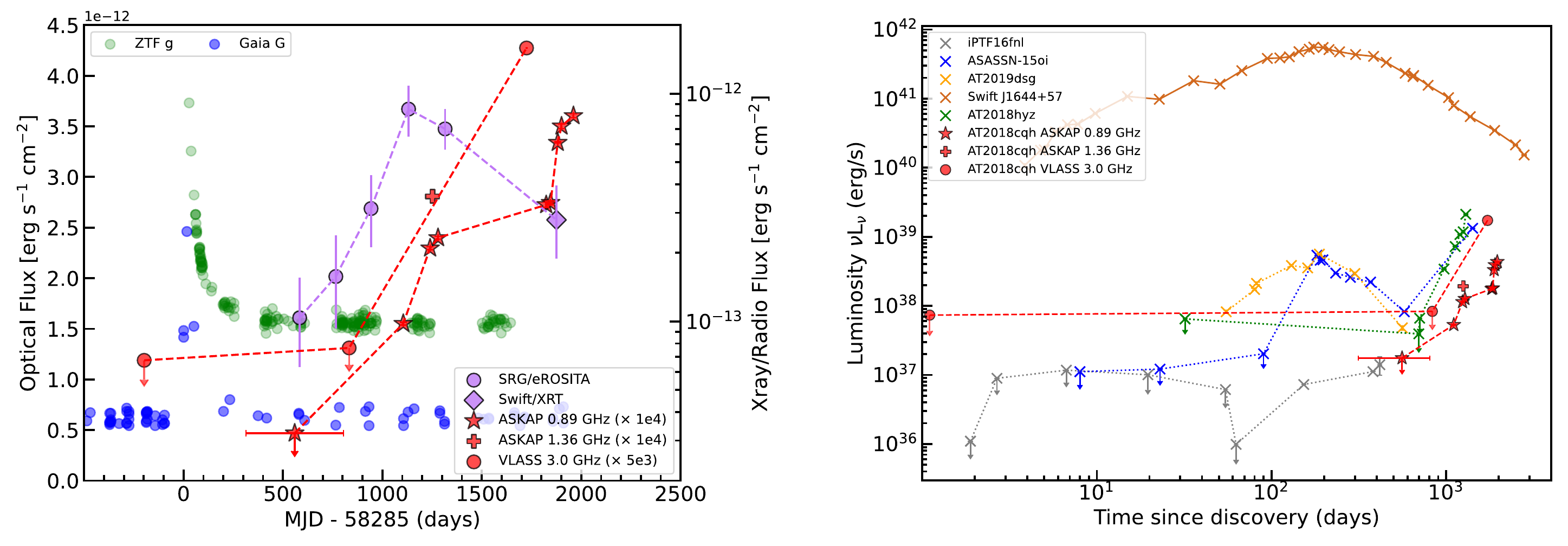}
 \hspace{0.1cm}
 \end{center}
  \vspace{-0.5cm}
\caption{
{\it Left Panel}: Light curves of \src in the optical (ZTF r-band and Gaia G-band), X-ray (0.3-2 keV), and radio (0.89, 1.36 and 3 GHz). Note that the optical light curves include the flux contributed by host. It is clear that there is a delayed brightening in the X-rays, while \src has faded in the optical to a quiescent level. The radio emission appears even later, for which the flux in still rising in both 0.89 GHz and 3 GHz. For the non-detections, the corresponding 3$\sigma$ upper limits on flux density are shown.  
{\it Right Panel}: The radio luminosity evolution of \src (red), including the upper limits. 
{Since the date of VLASS epoch I observations (2017 Nov 30) is before the time of optical discovery, it was shown for the purpose of 
illustrating the upper limit on the radio flux. }
Also shown for comparison are the light curves of the relativistic TDE Sw J1644+57 \citep[dark orange,][]{Cendes2021a}, and other three TDEs with late-rising radio emission: ASASSN-15oi 
\citep[blue;][]{Horesh2021a}, AT2018hyz \citep[green;][]{Cendes2022} and iPTF16fnl \citep[gray;][]{Horesh2021b}.
\label{fig:radiolc}}
 \vspace{0.2cm}
\end{figure*}

We searched for the radio emission from \src using the data from the Very Large Array Sky Survey \citep[VLASS,][]{Lacy2020}. 
VLASS is an on-going S-band (2–4 GHz) multi-epoch legacy survey aiming at to detect various types of extragalactic radio transients, 
including TDEs. 
VLASS observations are designed to survey the entire northern sky with Dec$>$ -$40^\circ$ (33,885 $\rm deg^2$) three times, 
with an angular resolution of 2$\farcs$5,
each separated by approximately a period of 32 months. 
Each VLASS epoch achieves an 1$\sigma$ sensitivity of $\sim$120 $\mu$Jy/beam, which is comparable to the depth of FIRST. 
The VLASS program has began in 2017, and recently completed its first and second epoch observations in 2019 (epoch I) 
and 2021 (epoch II), respectively.  
The epoch III observations are ongoing (from 2023 Jan to present). 
The preliminary ``QuickLook" images have been publicly released on the NRAO website\footnote{https://archive-new.nrao.edu/vlass/quicklook/}, 
in order to help the scientific community to timely access the VLASS data. 
While \src was not detected in both epoch I and II observations ($<$ 0.45 and 0.51 mJy/beam), we identify a bright radio transient in 
the recently released epoch III data (observed 2023 March 7), with a peak flux density of 10.58 $\pm$ 0.28 mJy/beam at 3 GHz, indicating a flux increase by a factor of $>$ 20.

The location of \src was also covered in the Australian Square Kilometre Array
Pathfinder (ASKAP) Variables and Slow Transients Survey \citep[VAST;][]{Murphy2021}, the Rapid ASKAP Continuum Survey \citep[RACS;][]{McConnell2020}, 
Evolutionary Map of the Universe Pilot Survey \citep[EMU;][]{Norris2021}, 
and the First Large Absorption Survey in H I \citep[FLASH;][]{Allison2022}.  
RACS is a large-area survey in the low frequency at 887.5 MHz covering the entire
radio sky south of declination $+41^{\circ}$ with typical sensitivities of 0.25-0.3 mJy/beam. 
RACS began observations in 2019 April and is now undertaking its mid-band observations at 1.36 GHz \citep{Duchesne2023}, 
with a resolution of $\sim$10\arcsec. 
By incorporating data from RACS, 
VAST was designed to detect astronomical phenomena that vary on timescales accessible in the ASKAP imaging mode (from
$\sim$5 s to several years). 
EMU Pilot Survey covers a total area of 270 square deg and is composed of 10 pointing fields, each of which was observed for 10 hrs at 943 MHz with an instantaneous 288 MHz bandwidth. 
The primary goal of EMU is to make a deep (rms noise levels of 10-20 $\mu$Jy/beam) radio continuum survey of the entire southern sky, extending as far north as +30$^\circ$.
FLASH is a wide field survey to detect 21-cm absorption lines in the continuum spectra of radio sources at intermediate redshifts, and is planned to cover 80\% of the sky at frequencies between 711.5 and 999.5 MHz. 
We retrieved all the RACS, VAST, EMU, and FLASH survey data currently available through the CASDA archive\footnote{https://data.csiro.au/domain/casda}, 
consisting of 20 individual images observed between 2019 Aug 27 and 2023 Oct 29, with a cadence down to 1 day. 
While there is no radio emission in the first 11 epochs, a bright radio source was detected after 2021 June 25 whose flux is still rising. 
This confirms the transient radio brightening of \src discovered by VLASS.

To secure the radio detections, we used the {\tt IMFIT} task in CASA to fit the radio emission component with a two-dimensional 
elliptical Gaussian model to determine the position, integrated and peak flux density. 
{The best-fit radio position from the VLASS epoch III observation is RA = $02^{\rm h}33^{\rm m}46\fs9361$ and DEC = -01$\arcdeg$01$\arcmin$28\farcs3760. 
In comparison with the position of optical flare given by ZTF (Section 2.1), we found a positional offset of $\sim$$0\farcs11$. 
This is a factor of three less than the astrometric accuracy of VLASS observations \citep[$\sim$$0\farcs4$,][]{Lacy2020}. 
Therefore, the optical and radio flares are spatially coincident, indicating that they are physically connected.  
}
For non-detections, we report the 3$\sigma$ upper limit on the flux, based on the map rms at the off-source position, which 
is in the range {249.3--616.0 $\mu$Jy/beam}. 
The radio emission at the three bands is unresolved and no extended emission is detected. 
The compactness of radio emission is confirmed by the ratios of integrated and peak flux density, 
which are in the range 0.95--1.23, with a median of 1.03. 
For consistency, only peak flux densities are used in our following analysis.  
The VLASS and ASKAP observation log and flux density measurements are presented in Table 1. 
Since the VAST, EMU and FLASH observations have little difference in the central frequency, we will adopt the same frequency of 0.89 GHz in our following analysis. 



\section{Analysis and results}

\subsection{Optical light curve analysis}

To characterize the optical light curve properties, we first fit a power-law model 
with the form: 

$$
L(t)=
\begin{cases} 
L_0\dfrac{t-t_D}{t_0-t_D} & t\leqslant t_0 \\
L_0\left(\dfrac{t-t_0+\tau}{\tau}\right)^{-p} & t>t_0 \\
\end{cases}
$$


where $L_0$ is the peak luminosity measured at the peak time $t_0$, $t_D$ represents the time of TDE, and $\tau$ is the {timescale} of luminosity decline. 
Note that only the data collected from the ZTF r-band and Gaia G-band observations are considered, as they have the same central wavelength, and the Gaia data detect an important rise to peak phase.  
The best-fit model results in a power-law index of $p=2.14_{-0.24}^{+0.31}$ and a time of disruption $t_D=-15_{-8.4}^{+4.7}$ days (relative to MJD 58285). 
Under this model the implied rise-time-to peak of $t_{\rm rise} = t_0 -t_D\approx37$ days. 
Fixing the decay index at the canonical value of $p=5/3$ for TDEs yields 
consistent results on the rise-time-to peak. 

 We further explored whether the optical light curves can be fitted with the Monte Carlo 
 software {\tt MOSFiT}, which has been applied to model the light curves of optical TDEs \citep{Mockler2019}. 
 The TDE model in {\tt MOSFiT} assumes that emission produced within an 
 elliptical accretion disk of a TDE is partly reprocessed into the UV/optical by an optically thick layer 
 \citep{Guillochon2018}. 
 We run {\tt MOSFiT} using a variant of the {\tt emcee} ensemble-based Markov Chain Monte Carlo routine
 until the fit has converged by reaching a potential scale reduction factor of $<$1.2 \citep{Mockler2019}. 
 In Figure \ref{fig:optlc}, we show 
 an ensemble of model realizations from {\tt MOSFiT}. The model is able to reproduce the data quite well, 
 including the stages of the rise to peak, near the peak, and the steady decline at later times. 
The best-fit model is that of a black hole of $2.5^{+0.7}_{-0.5}\times10^{6}$\msun~disrupting a {lower-mass} 
star of $\sim$$0.06\pm0.01$ \msun. 
 The stellar mass is similar to that inferred for other TDEs where a disrupted star with mass near 0.1\msun~is preferred \citep[][]{Mockler2019}. 
 Such a slight preference near 0.1\msun~ is likely due to the fact that below this mass the radius of the star is assumed constant in the {\tt MOSFiT} model, which favored for events in which short possible peak times are required \citep[][]{Mockler2019}. 

{With the best-fit {\tt MOSFiT} model, we obtained the monochromatic luminosity at the ZTF r-band of $\nu L_{\nu}\sim10^{43}$\erg, 
placing \src at the faint end of the luminosity function of optically-selected TDEs \citep{Lin2022, Yao2023}.   
The physical mechanisms that make a TDE faint remain poorly understood. 
Possible scenarios have been proposed including partial disruption, disruption of a lower-mass star \citep{Blagorodnova2017}, 
or complex outflow dynamics at low BH mass \citep{Charalampopoulos2023}. 
We note that the penetration parameter obtained from the {\tt MOSFiT} fittings is $\beta\equiv R_{\rm t}/R_{\rm p}=2.02^{+0.27}_{-0.20}$, 
where $R_{\rm t}$ refers to the tidal disruption radius and $R_{\rm p}$ is for the pericenter radius. 
The large $\beta$ ($>$1) indicates a deep encounter which is not compatible with a partial-disruption event. 
Therefore, the low luminosity could be attributed to a lower mass for the disrupted star and/or a lower-mass black hole, 
though the detailed underlying emission mechanism remains to be explored in future works. 
}

\subsection{Radio flux and SED evolution}

The radio light curves of \src at frequencies of 0.89, 1.36 and 3 GHz are shown in Figure \ref{fig:radiolc} (left panel). 
From the light curve at 0.89 GHz that is relatively well-sampled, we find a rapid rise at $t\simgt$1100 days since the optical discovery. 
At earlier epochs between 437 and 804 days, \src is observed by ASKAP but not detected. 
Stacking the individual ASKAP images results in a flux limit of $<$ 0.36 mJy/beam (at 3$\sigma$ level). 
The flux density rises by at least a factor of 25 from the non-detections at $t \sim 559$ days to a peak at $\sim$ 1961 days. 
The actual evolution of radio emission is characterized by a rise time of about 175 days, 
followed by a flattening for about 544 days, and a rebrightening at $\sim$ 137 days. 
This suggests a peculiar flux evolution with changing steep power-law ($F_{\rm \nu}\propto t^{\alpha}$) from index $\alpha\simgt$6.6, $\alpha\simgt$0.9 to $\alpha\simgt$ 29.0.  
Although the data are sparsely sampled, a similarly steep rise is observed at 3 GHz by VLASS from 833 (days) to 1725 (days), 
corresponding to $\alpha\simgt$ 4.2.

In Figure \ref{fig:radiolc} (right panel) we show the radio luminosity evolution of \srcs, as well as a comparison to previous radio-emitting TDEs, 
including jetted TDE Swift J1644+57 and those observed to exhibit late-time rise in the radio emission on timescales of 
hundreds of days. 
 The radio luminosity of \src increases from 1.8 $\times$ 10$^{37}$ erg s$^{-1}$ at 559 days to 1.7 $\times$ 10$^{39}$ erg s$^{-1}$ at 1725 days, making its evolution on a 
 {time scale} similar 
 to that of the TDE AT2018hyz \citep{Cendes2022} and secondary rising phase in the TDE ASASSN-15oi. 
 While the radio luminosity of AT2018hyz increases approximately linearly with time,  
 \src appears to display three phases in the radio evolution over the same epochs, with 
 a flattening between two rapid rising phases. 
 Due to the gap in the radio observations of \src between $\sim1280-1814$ days, 
 it is not clear whether the radio emission has declined in these epochs. 
Note that the overall luminosity evolution for \src is different from that of ASASSN-15oi, 
which had an initial bump in the light curve at $\sim180-550$ days. 
The flux limit of $<$ 0.36 mJy/beam at 0.89 GHz at earlier epochs between 437 and 804 days indicates that \src is not as luminous as ASASSN-15oi over the same epochs. 
On the other hand, the TDE iPTF16nl has also shown a radio re-brightening but at a much earlier epoch of $\approx$100 days \citep{Horesh2021b}. 
Given its more gradual rise in luminosity peaking at only $\sim$$10^{37}$\erg, the origin of radio emission in iPTF16nl might be different from that in \srcs.   
Finally, it should be noted that the radio luminosity of \src is
still at least one order of magnitude dimmer than that of Sw J1644+57 at a comparable timescale ($\approx$1100-2000 days), 
arguing against a process similar to that powering the radio emission in Sw J1644+57. 

\begin{figure}[ht!]
\epsscale{1.15}
\plotone{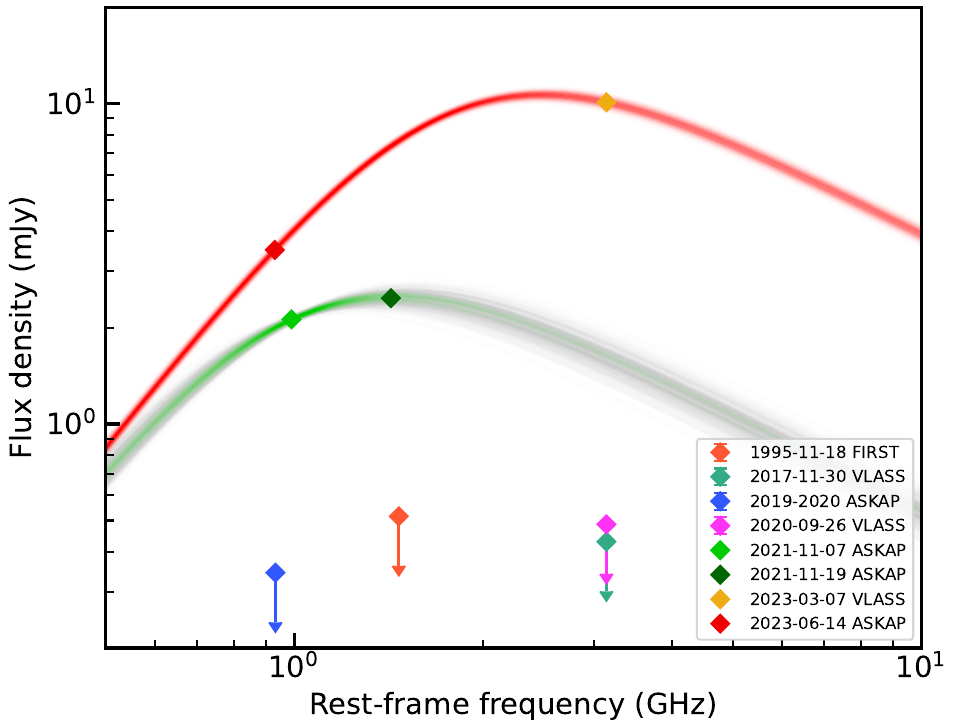}
\caption{The radio SEDs for two epochs which have quasi-simultaneous observations at different frequencies. For the non-detections, the corresponding 3$\sigma$ upper limits on flux density are shown. 
The red and green lines represent the best-fit to each SED from our MCMC modeling (Section 3.2), which are the model realizations on a basis of 500 random samples from the MCMC chains.  
There is a steadily rising in both peak flux density and frequency between the two epochs.  
}
\label{fig:radiosed}
\end{figure}

The multi-frequency radio data obtained for \src {allow} to model its spectral energy distribution (SED), 
which can in principle be used to 
constrain physical properties of radio-emitting region, 
such as an outflow expanding into and shocking the CNM \citep[e.g.,][]{Giannios2011, Metzger2012}. 
Figure \ref{fig:radiosed} shows the radio SEDs constructed using the data taking from quasi-simultaneously ASKAP and VLASS observations, at $t\sim 1246$ days and $t\sim 1725$ days, respectively. 
Note that we interpolated the flux at 0.89 GHz to $t\sim 1725$ days based on the observed radio light curve at between $t\sim 1814$ and $t\sim 1846$ days, to make it as quasi-simultaneous as possible to the flux at 3 GHz observed by VLASS epoch III. 
It is clear that the SED exhibits a gradual shift to a higher peak flux density and frequency, which is unprecedented in radio observations of TDEs \citep{Cendes2023}. 
We fit the SED evolution with the synchrotron emission models in the context of an outflow-CNM interaction, following the same approach outlined in \citep{Goodwin2022}. 
We assume no contribution to the transient radio emission from the host galaxy as it is not detected in pre-event archival FIRST observations 
(at least a factor of 5 fainter). 
The synchrotron emission spectrum is characterized by four parameters, namely 
$F_{0}$, $\nu_{m}$, $\nu_{a}$, and $p$, where $F_{0}$ is the flux normalization at 
$\nu_{m}$ (the synchrotron minimum frequency), $\nu_{a}$ is the synchrotron self-absorption frequency, and $p$ 
is the energy index of the power-law distribution of relativistic electrons.

\begin{figure*}[ht!]
\epsscale{1.15}
\plotone{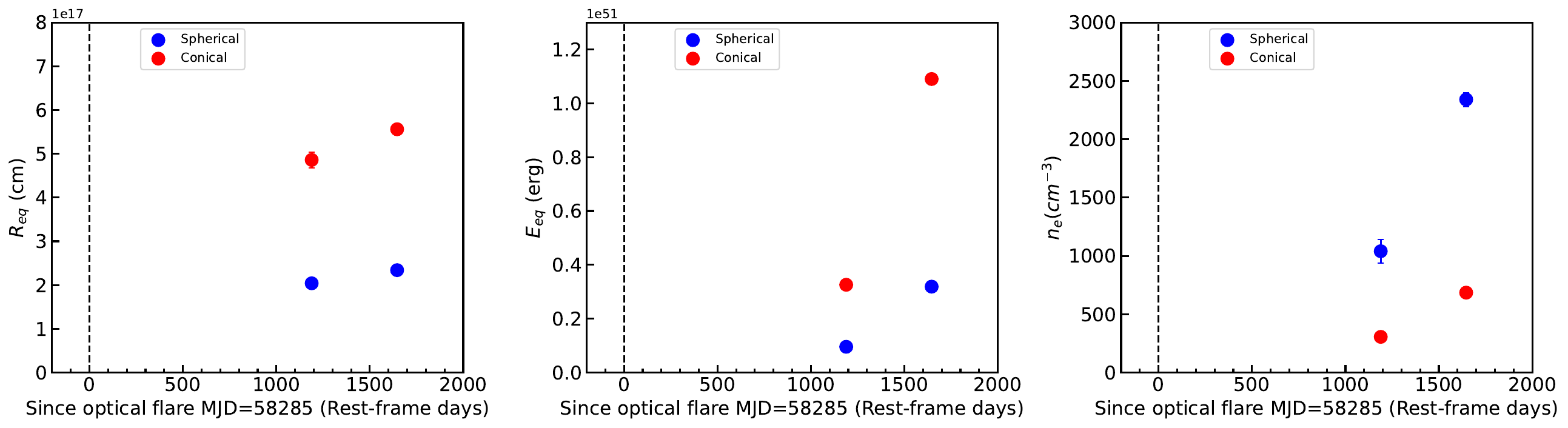}
\caption{The evolution of radius (left), kinetic energy (middle), and ambient density (right) as a function of time from our equipartition analysis assuming a single outflow is launched into the CNM around the time of optical discovery. Red and blue filled circles indicate parameters for a spherical homogeneous and a collimated, conical outflow, respectively. 
\label{fig:req}}
\end{figure*}

As in \citet{Goodwin2022},  
we use a Markov chain Monte Carlo (MCMC) fitting technique \citep[python module {\tt emcee},][]{Foreman-Mackey2013} 
to marginalize over the synchrotron model parameters to determine the best-fitting parameters and uncertainties. 
Due to the limited data points especially 
that at high frequencies ($>$3 GHz), we fix the synchrotron energy index 
to $p = 3$ \citep[e.g..][]{Alexander2016, Cendes2021a}. 
In fact, we find that the derived parameters do not deviate significantly within the $1\sigma$ uncertainties if adopting other reasonable values, such as $p \approx 2-3$.  
In Figure \ref{fig:radiosed}, we show the resulting SED models which provide a good fit to the data. 
From the SED fits we determine the peak flux density and frequency, $F_{\nu, p}$ and $\nu_{p}$, respectively.  
We find that both $F_{\nu, p}$ and $\nu_{p}$ indeed increase steadily with time, from 2.47 mJy and 1.42 GHz to 10.6 mJy and 2.48 GHz. 
Using the inferred values of $F_{\nu, p}$ and $\nu_{p}$, we can further assume equipartition to derive 
the radius of the radio emitting region ($R_{\rm eq}$) and kinetic energy ($E_{\rm eq}$)
using the scaling relations outlined in \citet{Barniol2013}. 
Following the procedures described in \citet{Goodwin2022}, we provide constraints for two different geometries, 
a spherical outflow and a mildly collimated conical outflow with a half-opening angle of $\phi=30^{\circ}$, in order 
to account for possible geometric dependence of outflow evolution. 

As shown in Figure \ref{fig:req}, assuming the spherical outflow, we find that the radius increases slightly from $R_{\rm eq}\approx$ 2.04 $\times$ 10$^{17}$ cm  to $\approx$ 2.34 $\times$ 10$^{17}$ cm between 1246 and 1725 days. 
The increase in $R_{\rm eq}$ becomes more rapidly for the case of a mildly collimated conical outflow. 
Under the assumption of free expansion, this corresponds to an outflow velocity {($v/c$)} of 0.024  (spherical) and 0.056 (conical) for the spherical (conical) geometries, with no sign of relativistic motion. 
Over the same epochs, the outflow kinetic energy increases by a factor of 3.3 (spherical), 
from $E_{\rm eq}\approx$ 9.57 $\times$ 10$^{49}$ erg to 3.19 $\times$ 10$^{50}$ erg. 
The kinetic energy is larger than that of other radio-emitting TDEs with non-relativistic outflows \citep{Cendes2023}. 
However, different from previous TDEs in which the ambient density profile is found 
approximately proportional to $R^{-2.5}$, 
we find that the inferred ambient density of CNM in \src increases with $R_{\rm eq}$ 
regardless of the outflow geometry. 
This may suggest that the outflow enters a high-density CNM structure. 
Because the radio light curve of \src is rising, indicating that the blast wave might be still in the expansion phase, 
continuous multi-frequency radio observations with a higher cadence 
will enable to better constrain the ambient density profile as a function of radius. 




\section{Discussion} 
\subsection{ A TDE from a candidate IMBH? }  

The host galaxy of \src 
is blue with an extinction-corrected rest-frame $u-r$ color $\approx$1.07. 
Bulge-disk decomposition of \src has been carried out by \citet{Simard2011}, 
which gives the bulge/total surface brightness ratio of $0.65\pm0.02$ at g-band, 
and a high S$\acute{e}$rsic index of $7.64\pm0.31$. 
The bulge stellar mass is estimated at log$(M_{\rm bulge}/{\rm\,M_\odot})=9.14^{+0.07}_{-0.05}$, compared to a total stellar mass of log$(M_{\rm stellar}/{\rm\,M_\odot})=9.51^{+0.16}_{-0.10}$ \citep{Mendel2014}. 
These measurements indicate a high central concentration
of stars, which has been seen within other TDE host galaxies \citep{French2020}. 
With the derived stellar mass, we can estimate the mass of the central BH to be 
log$(M_{\rm BH}/{\rm\,M_\odot})\approx5.88$ using a scaling region for 
low-mass galaxies \citep{Reines2015}. 

Spectral analysis suggests that only narrow emission lines are present in the pre-explosion 
SDSS spectrum after subtracting the host component. 
The ratios of the narrow lines place \src into the Seyfert (II) regime on the 
the Baldwin–Phillips–Terlevich (BPT) diagram (Figure \ref{fig:optspec}). 
{This may suggest that the multi-band flares of \src could be due to the AGN activity. 
However, this possibility seems unlikely based on the X-ray properties \citep{Bykov2023}. 
In addition, the rapid (duration of $<$1 yr) optical flare can also rule out the normal AGN variability 
as it is rarely seen in the light curves of AGNs \citep{Drake2011, Kankare2017, Zhang2022}. 
}
The lack of detectable broad permitted lines prevents us
from estimating the central BH mass of \src using the
conventional linewidth–luminosity–mass scaling relation. In the
optical spectral fittings, we measure the stellar velocity dispersion with Gaussian $\sigma_{\star} = 62\pm10$ 
km s$^{-1}$ after correcting for the instrumental broadening. Using
the $M_{\rm BH}-\sigma_{\star}$ relation for the low-mass end \citep{Xiao2011}, we estimated 
a BH mass of log$(M_{\rm BH}/{\rm\,M_\odot})\approx5.99$. 
With an intrinsic scatter of 0.5 dex, this is consistent with a relatively low BH mass 
of $M_{\rm BH}\approx5.9$ derived using the scaling relations between galaxy stellar mass and BH mass. 

The properties of the TDE light curve can potentially be used to
probe the BH mass\citep{Angus2022}, which could provide a measurement independent
of assumptions about the host galaxy. 
This is because the luminosity of TDE is expected to follow the fallback rate
of the stellar debris with a relationship \citep{Rees1988}:

  \begin{equation}
   t_{\rm fb}=41 M^{1/2}_{6}r_{*}^{3/2}m_{*}^{-1}\beta^{-3} \rm days
       \end{equation}
where $M_{6}$ is the BH mass in $10^{6}$\msun, $r_{*}$ and $m_{*}$ are the star's radius and mass in $R_\odot$ and $M_\odot$, 
and $\beta=R_{\rm T}/R_{\rm p}$ is the penetration factor \citep{Gezari2017, Bonnerot2020}. 
In Section 3.1, we have shown that when fitted the Gaia and ZTF r-band optical light
curve of \src using the {\tt MOSFit} code, a BH mass of log$(M_{\rm BH}/{\rm\,M_\odot})\approx6.42$ can be obtained. 
Because of the sparse sampling of the optical light curves during the peak phase, 
the actual rise to peak time may be even shorter, yielding a smaller BH mass.  
Combining with the BH masses estimated from the host galaxy properties, 
we can constrain the BH mass of \src in the range of $5.9<~$log$M_{\rm BH}/M_{\sun}<6.4$, placing it at the high mass end of the domain of IMBHs. 
Note that if the optical emission comes from process of stream-stream collision \citep{Piran2015}, rather than the accretion process hypothesized in {\tt MOSFit}, 
the correlation between the rise time of a TDE and the BH mass still holds as 
$t_{\rm fb}\propto M_{\rm BH}^{1/2}$. 
Therefore, although \src is a rare TDE whose host galaxy is one of only a small number of dwarf galaxies \citep[e.g.,][]{Angus2022}, we can not claim the presence of an IMBH with current data. 

\subsection{ Origin of the delayed X-ray flare} 
\src is one of few TDEs that has resolved rise-to-peak light curves in both X-ray and optical bands \citep{Gezari2017, Wevers2019, vanVelzen2020}. 
 The X-ray emission exhibits a delayed brightening roughly $\sim$ 590 days  
 with respect to the peak of optical emission which is also unique among optically discovered TDEs \citep{Guolo2023}. 
Many recent numerical studies have shown that the infalling stellar debris stream will 
 undergo self-intersections as a consequence of relativistic apsidal precession 
 \citep{Shiokawa2015, Bonnerot2017, Lu2020}, where 
 optical/UV emission could be produced because of shock heating. 
 Following the stream self-interactions, the debris spreads inward and gradually circularizes 
to form an accretion disk on the timescale of $\simeq(5-10)t_{0}$ \citep{Shiokawa2015, Bonnerot2016}, where $t_{\rm 0}$ is the orbital period of the most bound material (equation (1)).    
 {Within this picture, there will be a time delay between the debris self-crossing and onset of 
 disk formation, possibly explaining the observed delay of the X-ray emission in \srcs. 
For a BH mass of $\sim$$10^6$\msun, 
the circularization timescale 
can be estimated as $t_{\rm circ}\simeq200-400$ days, depending on the orbit eccentricity and penetration parameter \citep{Bonnerot2016}. 
In a more realistic scenario, if the viscous time ($t_{\rm visc}$) is not negligible (the time it takes material to accrete) and $t_{\rm visc}\simgt t_{\rm circ}$, the flare is prolonged at the expense of reduced peak luminosity \citep[e.g.,][]{Guillochon2015}. 
These predicted accretion properties seem not at odds with the long rise to peak time and relatively low peak luminosity ($L_{\rm 0.3-2 keV}\sim5.5\times10^{42}$\erg) 
observed with eROSITA \citep{Bykov2023}. 
 The blackbody radius inferred from the eROSITA observations 
 is $R_{\rm bb}\sim2\times10^{11}$ cm, comparable to the Schwarzschild radius ($R_{\rm s}=2GM_{\rm BH}/c^2$) 
 for a black hole mass of $\approx10^{6}$\msun. 
 This suggests that the soft X-ray emission indeed originates from a compact accretion disk. 

It has been proposed that if the majority of 
      falling-back debris becomes unbound in a dense outflow, the X-ray radiation from the inner accretion 
      disk will be initially blocked, and may escape at later times as the density and opacity of the 
      expanding outflow decreases \citet{Metzger2016, Wevers2019}. 
      In the model, efficient circularization of the returning debris is assumed, resulting in 
      rapid onset of disk accretion and reprocessed emission in the optical and UV bands. 
      However, the increase in the X-ray flux of \src lasts at least 550 days after the optical flare has decayed to the quiescent level. This is much longer than 
      the time scale for the ionization break out of X-ray radiation 
      for a black hole of $M_{\rm BH}\sim10^6$\msun~\citep{Metzger2016}. 
  In addition, assuming that the X-rays were produced at the time of optical flare, 
  we fitted the X-ray light curve using the data after the current peak ($t>$1000 days) with the canonical $t^{-5/3}$ decline law. Extrapolating the model backwards in time results in a unlikely peak luminosity of $>10^{50}$\erg, several orders of magnitudes above the Eddington luminosity. 
  On the other hand, the X-ray spectra are dominated by a soft blackbody component whose temperature remains little changed between the eROSITA observations, indicating no clear evidence for the decrease in absorbing column density with the increasing X-ray luminosity as expected in the reprocessing scenario. 
 Therefore, the scenario that the late time 
 X-ray brightening is due to the ionization break out of disk emission is disfavored .  


\subsection{ Origin of the delayed radio flare}  

Upon its radio detection, the steep rise in the flux density at 0.88 GHz (and possibly at 3 GHz as well) between 
$\approx 175$ days, $F_{\nu}\propto t^{\alpha}$ with $\alpha$ = 6.6,  
is not consistent with the theoretical predictions for an on-axis relativistic jet or a sub-relativistic outflow launched 
promptly after stellar disruption in \srcs. 
For instance, the fastest rise in the flux density is $F_{\nu}\propto t^{3}$ for an on-axis relativistic jet 
interacting with circumnuclear material (CNM) with a steep density profile of $\rho_{\rm CNM}\propto r^{-2.5}$ \citep{Horesh2021b}. 
Instead, to reconcile with the observed increase in flux density with the predictions of standard CNM shockwave models, 
a radio-emitting process that occurs at late times should be invoked.  
To achieve an $\sim t^{3}$ increase in flux density requires a delayed launch of the outflow by $\sim$ 600 days after optical discovery. 

In comparison with X-ray light curve, the evolution of the radio emission 
is clearly not paralleled, as the X-ray emission faded away after the peak, 
while the radio emission was still rising. 
In the context of TDEs, both theory and simulations suggest that the 
accretion rate of stellar debris onto a BH can vary by several orders of magnitudes \citep[e.g.,][]{Rees1988, Curd2019}. 
The combined X-ray and radio properties of \src appear to be reminiscent of the observed behavior of 
accretion state transition in X-ray binaries \citep[XRBs,][]{Fender2004}. 
In XRBs, a state transition in accretion occurs when the accretion
rate increases above a critical threshold, typically $\sim$$10-30$\% of the Eddington luminosity \citep{Done2007}. 
During the transition, the X-ray emission gradually becomes disk dominated. 
Given the peak X-ray luminosity of $L_{\rm 0.3-2 keV}\sim5.5\times10^{42}$\erg and BH mass $M_{\rm BH}\sim10^6$\msun, \src is likely {accreting at a rate of $L_{\rm Bol}/L_{\rm Edd}\sim0.1-0.5$ (depending on the bolometric correction factor). 
X-ray spectral analysis revealed that the X-ray spectra during the peak phase 
are dominated by a blackbody emission component (kT$_{\rm bb}\sim60$ eV), with a non-thermal hard tail that 
accounts for $\sim$10\% of the X-ray flux \citep{Bykov2023}. 
The non-thermal emission can be described by a powerlaw with $\Gamma=1.9$, 
likely from the low-luminosity AGN in the quiescent state (Section 4.1). 
}
It is therefore possible that a low-hard to high-soft phase transition occurred in \srcs,} 
resulting in the delayed launch of an outflow that led to rapidly rising radio emission at late times. 
Similar scenarios have been invoked to explain the late-time radio flares in the TDE ASASSN15-oi \citep{Horesh2021a} 
and AT2019azh \citep{Sfaradi2022}. 
However, standard accretion models can hardly explain the following evolution of radio flux after the initial steep rise, including a flattening lasting about 544 days and 
a phase with another steep rise.  


On the other hand, one may consider the possibility that the delayed radio emission from \src 
may be produced by a relativistic jet viewed off-axis, 
which is potentially relevant for TDEs in which the radio emission is still rising \citep{Matsumoto2023, Sfaradi2023}. 
In this model, the off-axis jet was initially launched at the time of TDE, 
which remains collimated with the emitting area increasing over time. 
The evolution of radio emitting region eventually intersects the light of sight to the observer, 
resulting in a delayed radio flare. 
On a basis of the best-fit synchrotron model to the radio SED evolution (Figure \ref{fig:radiosed}), 
we find both the peak flux densities and frequencies are increasing over time, 
with $F_{\rm peak}\propto t^{4.2}$ and $\nu_{\rm peak}\propto t^{1.8}$. 
Following the formalism of \citet{Matsumoto2023}, the apparent velocity of the radio emitting source 
is found to evolve with $\beta_{\rm eq, N}\propto t^{-0.8}$.  
Therefore, $\beta_{\rm eq, N}$ will continue to decrease monotonically unless either $F_{\rm peak}$ 
rises or $\nu_{\rm peak}$ decreases more rapidly, and the transition to the Newtonian branch will 
never happen, disfavoring the off-axis jet as the origin of delayed radio flares. 

Recently, \citet{Teboul2023} proposed a unified model for jet production in TDEs, 
which can be used to explain the delayed mildly-relativistic outflow observed in AT2018hyz \citep[see also][]{Lu2023b}. 
In the model, the late-time radio brightening in AT2018hyz  can be attributed to the break out of jet emission that was initially choked by the disk-wind ejecta, and its interaction with the CNM. 
However, once the decelerating jet expands into the CNM, the radio emission from the shock should follow a SED evolution similar to ASASSN-14li \citep{Alexander2016}, 
which is inconsistent with what is observed in \src (Figure 4). 
A similar scenario in which an outflow from the TDE interacts with 
dense clouds in an inhomogeneous CNM seems also less plausible. 
In this latter case, the peak frequency in the radio SED is predicted to decrease as a function of time, while the peak flux remains almost constant \citep{Bu2023}. 
Such a SED evolution was not observed in \srcs, at least with the current data. 
In addition, it would be challenging to explain the non-detection of infrared flares in the context of outflow-cloud interaction model. 
It should be noted that 
we have only quasi-simultaneous observations at two frequencies for a given epoch, 
so the current constraints on the SED evolution (hence $F_{\rm peak}$ and $\nu_{\rm peak}$) might not be robust. 
Since the radio flux is still rising, future radio observations covering a broader frequency range will be crucial 
to explore the exact evolution of peak flux densities and frequencies, allowing to better distinguish between different models in explaining the radio behaviour in \srcs. 



\section{Conclusion}
We present the discovery of delayed radio flare in an optical 
and X-ray detected TDE occurred in a dwarf galaxy. 
Both the optical light curve fitting and galaxy scaling relationships suggest a central black hole mass in the range of $5.9<~$log$M_{\rm BH}/M_{\sun}<6.4$. 
The temporal evolution of radio emission is peculiar, including an initial steep rise of 
at least 175 days, a flattening lasting about 544 days, and a phase with another steep rise.  
Although limited in the frequency coverage, the radio SED is found to evolve 
toward higher peak flux and frequencies over a period of $\approx$480 days.  
These properties make it challenging to explain the delayed radio brightening 
with an off-axis jet launched promptly after the TDE, the break out of a choked jet, 
or outflow-cloud interaction. 
The rapid rise in flux density coupled with the slow decay in the X-ray emission 
points to a delayed launching of outflow, perhaps due to a transition in the accretion state of the black hole.   
However, 
none of known accretion models can predict the radio variability behavior 
after the initial flare, including the flattening and secondary flare that is rising even more rapidly 
in comparison with the initial one. 
Since \src's radio emission is still rising, continued multi-frequency monitoring observations are required 
and crucial to understand the odd spectral and temporal properties of the delayed radio flares. 
\acknowledgments{
The data presented in this paper are based on observations
made with the Karl G. Jansky Very Large Array, the Australian SKA Pathfinder, the Zwicky Transient Facility, and the European Space Agency space mission Gaia. 
 The National Radio Astronomy Observatory is a facility of the
National Science Foundation operated under cooperative agreement
by Associated Universities, Inc.  
The Australian SKA Pathfinder is part of the Australia Telescope National Facility which is managed by CSIRO. Operation of ASKAP is funded by the Australian Government with support from the National Collaborative Research Infrastructure Strategy. 
This paper includes archived data obtained through the CSIRO ASKAP Science Data Archive, CASDA (http://data.csiro.au). 
The Zwicky Transient Facility Project is supported by the National Science Foundation under Grant No. AST-1440341. 
Gaia data are being processed by the Gaia Data Processing and Analysis Consortium (DPAC,
\url{https://www.cosmos.esa.int/web/gaia/dpac/consortium}). 
Funding for the DPAC
has been provided by national institutions, in particular the institutions
participating in the {\it Gaia} Multilateral Agreement.
We acknowledge ESA Gaia, DPAC and the Photometric Science Alerts Team (http://gsaweb.ast.cam.ac.uk/alerts), and the use of the Hale 200-inch Telescope through the Telescope Access Program (TAP), under 
the agreement between the National Astronomical Observatories, CAS, and the 
California Institute of Technology. 
 The work is supported by the SKA Fast Radio Burst
and High-Energy Transients Project (2022SKA0130102), and the National Science Foundation of China (NSFC) through grant No. 12192220 and 12192221. 
}

\software{CASA \citep[v5.3.0 and v5.6.1; ][]{McMullin2007}, 
MOSFiT \citep{Guillochon2017}, Astropy \citep{Astropy2013, Astropy2018, Astropy2022}
 }


\begin{thebibliography}{}

\bibitem[Alexander et al.(2016)]{Alexander2016} Alexander, K.~D., Berger, E., Guillochon, J., et al.\ 2016, \apjl, 819, L25.

\bibitem[Alexander et al.(2020)]{Alexander2020} Alexander, K.~D., van Velzen, S., Horesh, A., et al.\ 2020, \ssr, 216, 81. 

\bibitem[Allison et al.(2022)]{Allison2022} Allison, J.~R., Sadler, E.~M., Amaral, A.~D., et al.\ 2022, \pasa, 39, e010. 


\bibitem[Angus et al.(2022)]{Angus2022} Angus, C.~R., Baldassare, V.~F., Mockler, B., et al.\ 2022, Nature Astronomy, 6, 1452. 

\bibitem[Astropy Collaboration et al.(2013)]{Astropy2013} Astropy Collaboration, Robitaille, T.~P., Tollerud, E.~J., et al.\ 2013, \aap, 558, A33.

\bibitem[Astropy Collaboration et al.(2018)]{Astropy2018} Astropy Collaboration, Price-Whelan, A.~M., Sip{\H{o}}cz, B.~M., et al.\ 2018, \aj, 156, 123.

\bibitem[Astropy Collaboration et al.(2022)]{Astropy2022} Astropy Collaboration, Price-Whelan, A.~M., Lim, P.~L., et al.\ 2022, \apj, 935, 167.



\bibitem[Bade et al.(1996)]{Bade1996} Bade, N., Komossa, S., \& Dahlem, M.\ 1996, \aap, 309, L35


\bibitem[Barniol Duran et al.(2013)]{Barniol2013} Barniol Duran, R., Nakar, E., \& Piran, T.\ 2013, \apj, 772, 78.

\bibitem[Bellm et al.(2019)]{Bellm2019} Bellm, E.~C., Kulkarni, S.~R., Barlow, T., et al.\ 2019, \pasp, 131, 068003.

\bibitem[Blagorodnova et al.(2017)]{Blagorodnova2017} Blagorodnova, N., Gezari, S., Hung, T., et al.\ 2017, \apj, 844, 46

\bibitem[Bonnerot et al.(2016)]{Bonnerot2016} Bonnerot, C., Rossi, E.~M., Lodato, G., et al.\ 2016, \mnras, 455, 2253.

\bibitem[Bonnerot et al.(2017)]{Bonnerot2017} Bonnerot, C., Rossi, E.~M., \& Lodato, G.\ 2017, \mnras, 464, 2816.

\bibitem[Bonnerot \& Lu(2020)]{Bonnerot2020} Bonnerot, C. \& Lu, W.\ 2020, \mnras, 495, 1374. 

\bibitem[Bu et al.(2023)]{Bu2023} Bu, D.-F., Chen, L., Mou, G., et al.\ 2023, \mnras, 521, 4180. 

\bibitem[Bykov et al.(2024)]{Bykov2023} Bykov, S.~D., Gilfanov, M.~R., \& Sunyaev, R.~A.\ 2024, \mnras, 527, 1962. 

\bibitem[Cendes et al.(2021)]{Cendes2021a} Cendes, Y., Eftekhari, T., Berger, E., et al.\ 2021, \apj, 908, 125. 

\bibitem[Cendes et al.(2022)]{Cendes2022} Cendes, Y., Berger, E., Alexander, K.~D., et al.\ 2022, \apj, 938, 28. 

\bibitem[Cendes et al.(2023)]{Cendes2023} Cendes, Y., Berger, E., Alexander, K.~D., et al.\ 2023, arXiv:2308.13595. 

\bibitem[Charalampopoulos et al.(2023)]{Charalampopoulos2023} Charalampopoulos, P., Pursiainen, M., Leloudas, G., et al.\ 2023, \aap, 673, A95. 

\bibitem[Cordes \& Lazio(2002)]{Cordes2002} Cordes, J.~M. \& Lazio, T.~J.~W.\ 2002, arXiv:astro-ph/0207156

\bibitem[Curd \& Narayan(2019)]{Curd2019} Curd, B. \& Narayan, R.\ 2019, \mnras, 483, 565. 
  
\bibitem[Done et al.(2007)]{Done2007} Done, C., Gierli{\'n}ski, M., \& Kubota, A.\ 2007, \aapr, 15, 1

\bibitem[Drake et al.(2011)]{Drake2011} Drake, A.~J., Djorgovski, S.~G., Mahabal, A., et al.\ 2011, \apj, 735, 106. 

\bibitem[Duchesne et al.(2023)]{Duchesne2023} Duchesne, S.~W., Thomson, A.~J.~M., Pritchard, J., et al.\ 2023, \pasa, 40, e034. 

\bibitem[Fender et al.(2004)]{Fender2004} Fender, R.~P., Belloni, T.~M., \& Gallo, E.\ 2004, \mnras, 355, 1105

\bibitem[Foreman-Mackey et al.(2013)]{Foreman-Mackey2013} Foreman-Mackey, D., Hogg, D.~W., Lang, D., et al.\ 2013, \pasp, 125, 306.


\bibitem[French et al.(2020)]{French2020} French, K.~D., Wevers, T., Law-Smith, J., et al.\ 2020, \ssr, 216, 32. 

\bibitem[Gaia Collaboration et al. (2021)]{Gaia2021} Gaia Collaboration, Brown, A.~G.~A., Vallenari, A., et al.\ 2021, \aap, 649, A1

\bibitem[Gezari et al.(2017)]{Gezari2017} Gezari, S., Cenko, S.~B., \& Arcavi, I.\ 2017, \apjl, 851, L47.

\bibitem[Gezari(2021)]{Gezari2021} Gezari, S.\ 2021, \araa, 59, 21. 

\bibitem[Giannios \& Metzger(2011)]{Giannios2011} Giannios, D. \& Metzger, B.~D.\ 2011, \mnras, 416, 2102. 

\bibitem[Goodwin et al.(2022)]{Goodwin2022} Goodwin, A.~J., van Velzen, S., Miller-Jones, J.~C.~A., et al.\ 2022, \mnras, 511, 5328. 

\bibitem[Greene(2012)]{Greene2012} Greene, J.~E.\ 2012, Nature Communications, 3, 1304.

\bibitem[Guillochon \& Ramirez-Ruiz(2015)]{Guillochon2015} Guillochon, J. \& Ramirez-Ruiz, E.\ 2015, \apj, 809, 166. 

\bibitem[Guillochon et al.(2017)]{Guillochon2017} Guillochon, J., Nicholl, M., Villar, V.~A., et al.\ 2017, Astrophysics Source Code Library. ascl:1710.006

\bibitem[Guillochon et al.(2018)]{Guillochon2018} Guillochon, J., Nicholl, M., Villar, V.~A., et al.\ 2018, \apjs, 236, 6. 

\bibitem[Guolo et al.(2023)]{Guolo2023} Guolo, M., Gezari, S., Yao, Y., et al.\ 2023, arXiv:2308.13019. 

\bibitem[Hammerstein et al.(2023)]{Hammerstein2023} Hammerstein, E., van Velzen, S., Gezari, S., et al.\ 2023, \apj, 942, 9. 

\bibitem[Horesh et al.(2021a)]{Horesh2021a} Horesh, A., Cenko, S.~B., \& Arcavi, I.\ 2021a, Nature Astronomy, 5, 491. 

\bibitem[Horesh et al.(2021b)]{Horesh2021b} Horesh, A., Sfaradi, I., Fender, R., et al.\ 2021b, \apjl, 920, L5.

\bibitem[Kankare et al.(2017)]{Kankare2017} Kankare, E., Kotak, R., Mattila, S., et al.\ 2017, Nature Astronomy, 1, 865.

\bibitem[Lacy et al.(2020)]{Lacy2020} Lacy, M., Baum, S.~A., Chandler, C.~J., et al.\ 2020, \pasp, 132, 035001

\bibitem[Lin et al.(2022)]{Lin2022} Lin, Z., Jiang, N., Kong, X., et al.\ 2022, \apjl, 939, L33.

\bibitem[Lu \& Bonnerot(2020)]{Lu2020} Lu, W. \& Bonnerot, C.\ 2020, \mnras, 492, 686.

\bibitem[Lu \& Quataert(2023a)]{Lu2023} Lu, W. \& Quataert, E.\ 2023a, \mnras, 524, 6247.

\bibitem[Lu et al.(2023b)]{Lu2023b} Lu, W., Matsumoto, T., \& Matzner, C.~D.\ 2023b, arXiv:2310.15336.

\bibitem[Matsumoto \& Piran(2023)]{Matsumoto2023} Matsumoto, T. \& Piran, T.\ 2023, \mnras, 522, 4565. 

\bibitem[McConnell \& Ma(2013)]{McConnell2013} McConnell, N.~J. \& Ma, C.-P.\ 2013, \apj, 764, 184.

\bibitem[McConnell et al.(2020)]{McConnell2020} McConnell, D., Hale, C.~L., Lenc, E., et al.\ 2020, \pasa, 37, e048

\bibitem[McMullin et al.(2007)]{McMullin2007} McMullin, J.~P., Waters, B., Schiebel, D., et al.\ 2007, Astronomical Data Analysis Software and Systems XVI, 376, 127

\bibitem[Metzger et al.(2012)]{Metzger2012} Metzger, B.~D., Giannios, D., \& Mimica, P.\ 2012, \mnras, 420, 3528. 
  
\bibitem[Mendel et al.(2014)]{Mendel2014} Mendel, J.~T., Simard, L., Palmer, M., et al.\ 2014, \apjs, 210, 3. 

\bibitem[Metzger \& Stone(2016)]{Metzger2016} Metzger, B.~D. \& Stone, N.~C.\ 2016, \mnras, 461, 948.

\bibitem[Mockler et al.(2019)]{Mockler2019} Mockler, B., Guillochon, J., \& Ramirez-Ruiz, E.\ 2019, \apj, 872, 151. 

\bibitem[Mou \& Wang(2021)]
{mou2021}Mou, G., \& Wang, W., 2021, \mnras, 507, 1684

\bibitem[Mou et al.(2022)]{mou2022}Mou, G., Wang, T., Wang, W., \& Yang, J., 2022, \mnras, 510, 3650

\bibitem[Murphy et al.(2021)]{Murphy2021} Murphy, T., Kaplan, D.~L., Stewart, A.~J., et al.\ 2021, \pasa, 38, e054. 

\bibitem[Nakar \& Granot(2007)]{Nakar2007} Nakar, E. \& Granot, J.\ 2007, \mnras, 380, 1744. 

\bibitem[Norris et al.(2021)]{Norris2021} Norris, R.~P., Marvil, J., Collier, J.~D., et al.\ 2021, \pasa, 38, e046. 

\bibitem[Oke \& Gunn(1982)]{Oke1982} Oke, J.~B. \& Gunn, J.~E.\ 1982, \pasp, 94, 586. 

\bibitem[Piran et al.(2015)]{Piran2015} Piran, T., Svirski, G., Krolik, J., et al.\ 2015, \apj, 806, 164. 

\bibitem[Prochaska et al.(2020a)]{Pypeit1} Prochaska, J., Hennawi, J., Westfall, K., et al.\ 2020, The Journal of Open Source Software, 5, 2308. 

\bibitem[Prochaska et al.(2020b)]{Pypeit2} Prochaska, J.~X., Hennawi, J., Cooke, R., et al.\ 2020, Zenodo

\bibitem[Rees(1988)]{Rees1988} Rees, M.~J.\ 1988, \nat, 333, 523. 

\bibitem[Reines et al.(2013)]{Reines2013} Reines, A.~E., Greene, J.~E., \& Geha, M.\ 2013, \apj, 775, 116. 

\bibitem[Reines \& Volonteri(2015)]{Reines2015} Reines, A.~E. \& Volonteri, M.\ 2015, \apj, 813, 82. 

\bibitem[Reines(2022)]{Reines2022} Reines, A.~E.\ 2022, Nature Astronomy, 6, 26. 

\bibitem[Saxton et al.(2020)]{Saxton2020} Saxton, R., Komossa, S., Auchettl, K., et al.\ 2020, \ssr, 216, 85


\bibitem[Sexton et al.(2021)]{Sexton2021} Sexton, R.~O., Matzko, W., Darden, N., et al.\ 2021, \mnras, 500, 2871.

\bibitem[Sfaradi et al.(2022)]{Sfaradi2022} Sfaradi, I., Horesh, A., Fender, R., et al.\ 2022, \apj, 933, 176. 

\bibitem[Sfaradi et al.(2023)]{Sfaradi2023} Sfaradi, I., Beniamini, P., Horesh, A., et al.\ 2023, arXiv:2308.01965. 

\bibitem[Shiokawa et al.(2015)]{Shiokawa2015} Shiokawa, H., Krolik, J.~H., Cheng, R.~M., et al.\ 2015, \apj, 804, 85.

\bibitem[Shu et al.(2020)]{Shu2020} Shu, X., Zhang, W., Li, S., et al.\ 2020, Nature Communications, 11, 5876. 

\bibitem[Simard et al.(2011)]{Simard2011} Simard, L., Mendel, J.~T., Patton, D.~R., et al.\ 2011, \apjs, 196, 11. 

\bibitem[Stone et al.(2019)]{Stone2019} Stone, N.~C., Kesden, M., Cheng, R.~M., et al.\ 2019, General Relativity and Gravitation, 51, 30. 

\bibitem[Teboul \& Metzger(2023)]{Teboul2023} Teboul, O. \& Metzger, B.~D.\ 2023, \apjl, 957, L9. 

\bibitem[van Velzen et al.(2020)]{vanVelzen2020} van Velzen, S., Holoien, T.~W.-S., Onori, F., et al.\ 2020, \ssr, 216, 124.

\bibitem[van Velzen et al.(2021)]{vanVelzen2021} van Velzen, S., Gezari, S., Hammerstein, E., et al.\ 2021, \apj, 908, 4. 

\bibitem[Wevers et al.(2019)]{Wevers2019} Wevers, T., Stone, N.~C., van Velzen, S., et al.\ 2019, \mnras, 487, 4136. 

\bibitem[Xiao et al.(2011)]{Xiao2011} Xiao, T., Barth, A.~J., Greene, J.~E., et al.\ 2011, \apj, 739, 28. 

\bibitem[Yao et al.(2023)]{Yao2023} Yao, Y., Ravi, V., Gezari, S., et al.\ 2023, \apjl, 955, L6.

\bibitem[Zhang et al.(2022)]{Zhang2022} Zhang, W.~J., Shu, X.~W., Sheng, Z.~F., et al.\ 2022, \aap, 660, A119. 

\end{thebibliography}
\end{document}